# QUANTUM KERNEL FOR IMAGE CLASSIFICATION OF REAL WORLD MANUFACTURING DEFECTS

**Daniel Beaulieu**

Deloitte Consulting LLP
Arlington, VA 22209
dabeaulieu@deloitte.com

**Dylan Miracle**

Strangeworks Inc.
Austin, TX 78702
dylan@strangeworks.com

**Anh Pham**

Deloitte Consulting LLP
Atlanta, GA 30303
anhdpham@deloitte.com

**William Scherr**

Deloitte Consulting
LLP Arlington, VA
22209
wscherr@deloitte.com

December 16, 2022

# 1 Abstract

The quantum kernel method results clearly outperformed a classical SVM when analyzing low-resolution images with minimal feature selection on the quantum simulator, with inconsistent results when run on an actual quantum processor. We chose to use an existing quantum kernel method for classification. Quantum computers can efficiently find similarities between data points even for nonlinear problems by mapping them to higher dimensional quantum Hilbert spaces in order to better separate data. Quantum Kernel methods are theorized to outperform classical support vector machines (SVMs) when the data has non-linear relationships or has a large and complex feature space. We applied dynamic decoupling error mitigation using the Mitiq package to the Quantum SVM kernel method, which, to our knowledge, has never been done for quantum kernel methods for image classification.

We applied the quantum kernel method to classify real world image data from a manufacturing facility using a superconducting quantum computer. The manufacturing data used was quality assurance images used to determine if a product was defective or was produced correctly through the manufacturing process. We also tested the Mitiq dynamical decoupling (DD) methodology to understand effectiveness in decreasing noise-related errors.

We also found that the way classical data was encoded onto qubits in quantum states affected our results. All three quantum processing unit (QPU) runs of our angle encoded circuit returned different results, with one run having better than classical results, one run having equivalent to classical results, and a run with worse than classical results. The more complex instantaneous quantum polynomial (IQP) encoding approach showed better precision than classical SVM results when run on a QPU but had a worse recall and F1-score. We found that Mitiq DD error mitigation     did not improve the results of IQP encoded circuits runs and did not have an impact on angle encoded circuits runs on the QPU. In summary, we found that the angle encoded circuit performed the best of the quantum kernel encoding methods on real quantum hardware, and a non-error corrected IQP circuit performed second best for detecting errors in manufacturing image data. In future research projects using quantum kernels to classify images, we recommend exploring other error mitigation techniques than Mitiq DD.

## 2 Introduction and Motivation

### 2.1 Quantum Kernel Methods

Machine learning and AI algorithms are considered a prime candidate for near term quantum advantage relative to other fields in quantum computing[5]. Quantum kernels have been found to be able to more accurately perform a supervised classification task for difficult to classify compared to a classical SVM[8]. The idea of using kernel methods to take advantage of the larger problem space offered by quantum Hilbert space to separate two-dimensional data into a higher-dimensional space is a relatively new concept[4]. Havlicek et al. 2018 proposed the concept of using a quantum computer to estimate the kernel function of the quantum Hilbert feature space and run a classical SVM using the quantum kernel created on the QPU[4]. They argued that quantum kernels methods are ideal for noisy intermediate scale quantum computers since it is difficult for a classical SVM to reproduce the output distribution as it is extremely difficult to classically simulate a problem space as vast as quantum Hilbert space for a IQP circuit with multiple layers[4].

Since then, many other authors have progressed the state of the art for quantum kernels. In working with synthetic weather radar[1], Enos et al. 2021found that quantum kernels could, in principle, perform fundamentally more complex tasks than classical learning machines on the relevant underlying data[5]. Peters et al. 2021 was able to classify a data set of astronomical events to classify supernova with no dimensionality reduction, and their results achieved accuracy that is comparable to classical techniques[11]. Current research draws from theorems put forth by Schuld 2021, who argues that quantum machine learning should be done using quantum kernel methods since popular quantum machine learning methods map data to a high-dimensional quantum Hilbert space, and that in the NISQ era, quantum kernel methods should find better or equally accurate findings as variation circuit based and require less circuit depth[14]. Research has also found that quantum kernel alignment can make use of group structure to improve model predictions[2]. This project was intended to test how quantum kernel methods compare to classical machine learning support vector machine (SVM) techniques when classifying data from a real-world business problems. While the Liu et al. 2020 paper uses synthetic data sets intentionally constructed to be difficult for classical SVM methods to classify, the current research seeks to apply a quantum kernel method to a real-world classification task using actual data from a manufacturing facility. The largest contribution to research on applying Quantum Kernel methods is the inclusion of Mitiq error mitigation. In addition, our research contributes to the body of research on image classification by using a different pre-processing technique than Guijo et al. 2022[3] for manufacturing image and applying quantum kernel SVMs for classifying manufacturing defects on real-world data.

### 2.2 Goals:

The primary goal is to test the efficacy of different quantum kernel implementations compared to classical implementations. The secondary goal is to test an IQP encoding method of the quantum kernel vs a simple angle encoded quantum kernel to see if there is improvement in performance on actual NISQ quantum hardware. In addition, due to the complexity of the IQP circuit encoding method, our research will explore if error mitigation can potentially improve the performance on the quantum hardware using the Mitiq DD package developed by the Unitary fund[7].

## 3 Data

The data used for this research is images of housings for Smart Rover STEM educational kits manufactured at The Smart Factory @ Wichita in partnership with Elenco Electronics. The Smart Factory @ Wichita is a Deloitte-led ecosystem of innovative collaborators with a shared goal: demonstrating how an authentic smart factory is a holistic endeavor, moving beyond the shop floor, to improve efficiency, productivity, and sustainability across the greater enterprise. The data used in this research was created from a collaboration with The Smart Factory @ Wichita to help improve manufacturing of Smart Rover modules, which contains a Raspberry Pi microcomputer, camera, body housings, and other accessories. The images analyzed by the models are from the collection of Smart Rover body housings which may include varying grades of defects and imperfections to the body housing.

The raw image data was encoded using the sci-kit learn and Imageio python packages[11]]. Our preprocessing technique removed color from the images, rendering them in grey scale. The image size was shrunk down to 28x28. The features in the image were further reduced using Principal Components Analysis (PCA) so that the number of columns in the data set would be equal to the number of qubits so the problem would fit on the QPU. Principal components analysis is a widely used dimensionality reduction technique used in data science applications. For more information about the PCA technique please consult Shlens' 2014 tutorial[17].

## 4 Methods

### 4.1 Support Vector Machines

A support vector machine (SVM) is a supervised learning algorithm which can be used for both classification and regression problems. An SVM finds a hyperplane that separates data points of one class from those of another class with the largest margin between the two classes. The "margin" is the maximal width of the area parallel to the hyperplane that has no data points within it. Hyperplanes can only be found for linearly separable problems. SVMs can utilize multiple types of kernels which can allow them to be more flexible and handle nonlinear problems[13].

### 4.2 Quantum Kernel Method

The method used to create the quantum kernel in this work is based on Enos et al. 2021 formulation of the quantum kernel described in Havlicek et al. 2018 where a kernel is estimated on a quantum computer for all pairs of training data. The creation of a quantum kernel is through a process where data is first encoded to a feature map, then calculated as an inner product between two data-encoding feature vectors, then finally encoded as a scalar input into the quantum state of a single qubit[6]. This allows us to linearly separate data in a high-dimensional Hilbert space that would be difficult to separate linearly using classical methods. The number of qubits required by our model is equal to the number of features in the data set, thus the training data set is a matrix of the number of training samples times the number of features/qubits which gets mapped to a quantum Hilbert feature space. As a result, we used PCA to reduce the size of the of the data set so that the number of features equaled the number of qubits.

To create a quantum feature map, we encoded the Kernel represented in equation 1.

$$\widetilde{m}(\vec{s}) = sign\left(\sum_{i=1}^{t} y_i \alpha_i^* K(\vec{x}_i, \vec{s}) + b\right) \quad (1)$$

The kernel used in the classical SVM, $K(\vec{x}_i, \vec{x}_j)$, is created by examining all pairs of training data on a QPU. Then the optimization problem that will be solved by the classical SVM is created using a dual quadratic program which accesses this quantum generated kernel[4].

$$L_D(\alpha) = \sum_{i=1}^{t} \alpha_i - \frac{1}{2}\sum_{i,j=1}^{t} y_i y_j \alpha_i \alpha_j K(\vec{x}_i, \vec{x}_j) \quad (2)$$

### 4.3 Data Encoding

To get the classical data to run on a quantum computer with our quantum kernel classifier, we first had to encode the data into a quantum state that can be represented by qubits. As a baseline encoding methodology, we chose the angle encoding method which represents classical data through applying gate rotations around the X axis. The Angle encoding method is efficient in terms of operations but does not allow for entanglement[12]. Alternatively, we implemented the Instantaneous Quantum Polynomial (IQP) encoding strategy. IQP is more complex and encodes classical data as a unitary that applies repeated layers of Hadamard gates upon all qubits in parallel[6] along with CPHASE gates. IQP including entanglement allows us to take advantage of the quantum-mechanical properties of qubits, exponentially expanding the calculation capacity of QPUs as qubit counts increase[15].

Angle Encoded Circuit

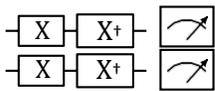

IQP Encoded Circuit

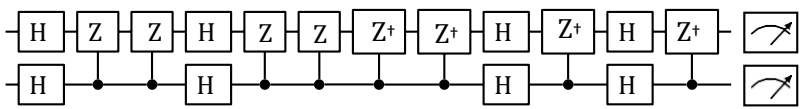

### 4.4 Error Mitigation Strategies

In order to improve our results through error mitigation, we used the Mitiq package from the Unitary Fund[7]. Mitiq is a Python package which includes multiple types of error mitigation techniques for a variety of quantum computers. The error mitigation technique we used is dynamical decoupling (DD)[9],[10],[18], which helps mitigate error by decoupling qubits from environmental noise through a rapid sequences of Pauli control pulses. The DD method implemented in Mitiq adds different series of Clifford gates like XX or XYXY between the two-qubit controlled gates like CPHASE. We focused on DD due to the ease of integration with the PyQuil SDK. As a result, the circuit depth for IQP encoding method with DD error mitigation was increased with the DD error mitigation technique, while the DD technique did not modify the circuit of angle encoding. It is important to note that angle encoding method does not include any two-qubit controlled gates.

### 4.5 Evaluation Metrics

From the Scikit-learn documentation "Precision is the ratio $tp/(tp + fp)$ where tp is the number of true positives and fp the number of false positives. The precision is intuitively the ability of the classifier not to label a negative sample as positive.

The recall is the ratio $tp/(tp + fn)$ where tp is the number of true positives and fn the number of false negatives. The recall is intuitively the ability of the classifier to find all the positive samples.

The F1 score can be interpreted as a weighted harmonic mean of the precision and recall, where an F-beta score reaches its best value at 1 and worst score at 0. The relative contribution of precision and recall to the F1 score are equal[16].

### 4.6 Study Limitations

We found that we could not simulate circuits consisting of more than 20 qubits due to the constraints of the simulator. Thus, we used PCA to reduce the features of the data to 20 qubits from the 28 columns of image data per image. We could have run the model for the QPU on larger images but would not have been able to benchmark our results against the simulator. To get the largest number of simulated qubits, we used the Wavefunction simulator without a noise function, which may have impacted the direction of our research away from potentially promising avenues of research.

Due to budgetary constraints and limited access to the QPU, we limited the number of runs and could not investigate every combination of parameters to test different error mitigation techniques implemented in the Mitiq (with different encoding types). Therefore, we did not use the K-crossfolds validation technique nor run multiple random samples to ensure our sample selection was not an outlier that provided unique results. We also had to accept we had a high level of inconsistency with the three-angle encoded runs.

We evaluated different quantum hardware, including quantum annealers and gate-based technologies, and decided to perform our classification tasks using a quantum kernel method on the 80-qubit Rigetti Aspen-M-2 QPU since it was the largest QPU we had access to    .

We were unable to use the QPU's active reset feature on our qubits because the Mitiq package is not compatible with this feature. Deactivating active reset may have impacted our results when using Mitiq for some DD runs. The simulated results returned identical results for IQP and Angle encoding.

## 5 Technology Stack

The technology stack used for this work is Strangeworks platform, which is a cloud hosted data science platform which integrates on the back end with a wide variety of quantum hardware. Strangeworks provides access to Rigetti QPUs by wrapping PyQuil functions such as run and compile into a Python SDK that is imported into the user's code. The SDK connects job requests to QPU hardware as well as provides job tracking through the SDK and the Strangeworks web portal. The platform intermediates the data exchange between the user and the backends required for running circuits on the QPU. These include a hosted Quilc compiler, the Quil compiler for QPUs, as well as a hosted quantum virtual machine (QVM). While Strangeworks itself does not perform any additional transpilation or optimization to experiments, hosting their services near the quantum resources decreases latency between the compiler, the virtual machine, and the QPU. The impact of round-trip latency between the user and external systems is the largest contributor to overall experiment completion time.

For this research project, the code was written in the PyQuil quantum computing Python package and run on the Rigetti Aspen-M-2 QPU. The code was compiled using the Quilc compiler. In addition, the results were also simulated on the Wavefunction Simulator to compare with the hardware run.

## 6 Results

The results on the simulated quantum kernel runs showed improvement over the classical SVM kernel using a linear kernel for manufacturing images for angle and IQP encoded circuits when run on the simulator. This indicates the quantum kernel method can perform well in noiseless simulation. Comparisons against other classical SVM kernels are shown in the appendix.

Our angle encoded results for the manufacturing images on QPU shows that we achieve better than classical results for one of the three runs, equivalent results for a second run, but significantly worse results with one of the three runs. You can see the different angle encoded runs on QPU on Table 1 below. The IQP results performing more consistently than Angle encoding on QPU, but worse than the classical SVM methods. Mitiq did not change the angle encoded results and did not have a positive impact upon IQP encoded results in terms of precision, recall, and F1-score. The IQP results with Mitiq DD error mitigation were worse than the non-error encoded results. As shown in the results, the choice of Mitiq DD sequence did not impact performance, with XX and XYXY performing worse than the classical linear kernel SVM for IQP encoded circuits. We were not able to test the Mitiq DD YY Pauli sequence, zero noise estimation, or probabilistic error mitigation techniques due to limited QPU access.

Table 1: Quantum Kernel Angle Encoding Results with Image Data (N=60, Train=42, Test=18)

|  | Non QPU | | QPU | | | |
| --- | --- | --- | --- | --- | --- | --- |
|  | Classical | Simulated | Angle Run 1 | Angle Run 2 | Angle Run 3 | Average Angle Run |
| Precision | 0.68 | 0.73 | 0.25 | 0.75 | 0.67 | 0.56 |
| Recall | 0.67 | 0.72 | 0.5 | 0.72 | 0.67 | 0.63 |
| F1-score | 0.66 | 0.72 | 0.33 | 0.71 | 0.67 | 0.57 |
| Run Time: CPU/QPU | 2s | 3min, 55s | 32.7s | 28.3s | 29.1s |  |
| Run Time: Wall | 2s | 45min, 36s | 7min, 44s | 28min, 54s | 28min, 45s |  |

Table 2: Quantum Kernel IQP Encoding Results with Image Data (N=60, Train=42, Test=18)

|  | Non-QPU | | QPU | | |
| --- | --- | --- | --- | --- | --- |
|  | Classical | Simulated | IQP No Mitiq | IQP Mitiq DDD XYXY | IQP Mitiq DDD XX |
| Encoding | N/A | IQP | IQP | IQP | IQP |
| Mitiq Implementation | N/A | N/A | No | Yes | Yes |
| Correction Type | N/A | N/A | N/A | XYXY | XX |
| Precision | 0.68 | 0.73 | 0.78 | 0.25 | 0.25 |
| Recall | 0.67 | 0.72 | 0.61 | 0.50 | 0.50 |
| F1-score | 0.66 | 0.72 | 0.54 | 0.33 | 0.33 |
| Run Time: CPU/QPU | 2s | 3min, 55s | 35.6s | 36.4s | 38.5s |
| Run Time: Wall | 2s | 45mins, 36s | 12min | 32min, 5s | 33min, 53s |

### 6.1 Comparison of Classical Kernels for Manufacturing Image Data

Table 3 below shows how the Mitiq DD did not improve results for IQP encoded circuits and actually made predictions in the quantum kernel SVM worse. However, we did have two of three angled encoded runs that performed as well or better than all classical SVM methods, including Radial Basis Function (RBF), Polynomial, Linear, and Sigmoid kernels in terms of precision and recall. The F1-score for RBF is slightly higher than the best quantum angle encoded runs, but both precision and recall are worse.

As you can see in Figure 5.1 the first QPU run of our angle encoded circuit performed poorly compared to the classical linear SVM model, while the second QPU angle encoded run performed slightly better, and the third QPU run performed equivalently.

## 7 Discussion and Conclusions

As demonstrated in our results, the quantum kernel generated on a QPU performed better than a linear kernel on a classical SVM. The best performing circuit is angle encoded with no error mitigation applied. The structured classical data sets all performed more poorly on actual quantum hardware than the unstructured raw image files we analyzed.

We believe the pre-processing technique used to transform Smart Factory images down to grey scale 28x28 and matching the number of qubits, may have created an input training matrix for the SVM that was more conducive to performing well on the QPU. When we used a 200x200 image size and shrunk it down to 20 qubits, the simulator returned much worse results. This is similar to what Guajo et al. 2022 did to pre-process images of manufacturing defects, though we shrunk the image first prior to the PCA process[3]. The advantage of our method compared to Guijo et al. 2022 is that PCA allowed us to intelligently shrink our data dimensions while retaining as much information as possible and encode that data onto the number of qubits we were able to use.

We believe that the reason angle encoding quantum kernel circuits performed raw image classification tasks equivalent to that of classical SVM when the data dimensionality is reduced to the number of qubits using PCA is that quantum kernel methods can efficiently encode highly non-linear data by using a feature map. Specifically, this feature map allows us to utilize the higher dimensionality of quantum Hilbert problem space relative to other kernel types [4]. However, it seems that even in the absence of two-qubit controlled gates the angle encoding circuits are also highly influenced by decoherent noise, thus resulting in inconsistent results in different runs. Further, we believe that the increased length of IQP circuits and their use of two-qubit gates made performance consistently worse on QPU than simulator. Specifically, as shown in Tabled 1 and 2, the results on the simulator for both angle and IQP encoded circuits show better performance of the quantum kernel technique than the classical SVM results. All good images have to be nearly identical to work, while images of defective manufacturing products can be defective in nearly any way. Our interpretation is that the quantum kernel method was able to take advantage of non-linearities in our image data set more efficiently in our processed images better than a classical linear kernel support vector machines. We are planning a future study using medical images which are reduced to small dimensions while retaining important details.

We also believe that, even though some angle-encoded circuit runs outperformed the classical linear SVM method on the simulator, angle encoding was highly inconsistent across runs on actual quantum hardware. Implementing Mitiq's DD error mitigation lengthens the IQP circuits, and on current noisy intermediate-scale quantum hardware this may have reduced performance compared to non-error corrected results from implementing Mitiq's dynamic decoupling technique. We recommend that any production-oriented quantum classification processes, including classifying manufacturing defects like in this research, experiment with both angle and IQP and experiment with different methods of error mitigation since both performed better in simulated quantum kernel results. Improving the consistency of our results with angle encoding would improve the deployment of quantum machine learning models for useful business data science tasks.

Key take away lessons from our research:

1. Conversion of raw image data into a matrix format with a small image size and using PCA to reduce the number of features to the number of qubits, performed better than a classical SVM on simulator.

2. The noiseless simulated results for both angle and IQP encoded circuits classified images better than any classical SVM kernels, indicating our quantum kernel method is effective, but may be affected by error when running on quantum hardware

3. Our angle encoded results performed inconsistently, with one run being better, one run performing equivalently, and one run performing significantly worse than classical SVM kernels.

4. IQP encoded circuits consistently underperformed on QPU runs than classical linear SVM and did not benefit from Mitiq DD error mitigation for our manufacturing defect image classification problem.

5. Dynamical decoupling as implemented in Mitiq had no effect on angle encoding. We recommend experimenting with different types of error mitigation to improve consistency and reduce errors.

Table 3: Comparison of Linear Kernel Methods with Manufacturing Image Data

| Kernel | Precision | Recall | F1-Score |
|---|---|---|---|
| **Classical SVM Methods** | | | |
| Linear | 0.68 | 0.67 | 0.66 |
| Poly | 0.5 | 0.5 | 0.41 |
| RBF | 0.76 | 0.56 | 0.45 |
| Sigmoid | 0.40 | 0.44 | 0.38 |
| **Quantum Kernels** | | | |
| Angle Run 1 | 0.25 | 0.50 | 0.33 |
| Angle Run 2 | 0.75 | 0.72 | 0.71 |
| Angle Run 3 | 0.67 | 0.67 | 0.67 |
| IQP No Mitiq | 0.76 | 0.56 | 0.45 |
| IQP W/ Mitiq (XYXY) | 0.32 | 0.33 | 0.32 |
| IQP W/ Mitiq (XX) | 0.32 | 0.33 | 0.32 |

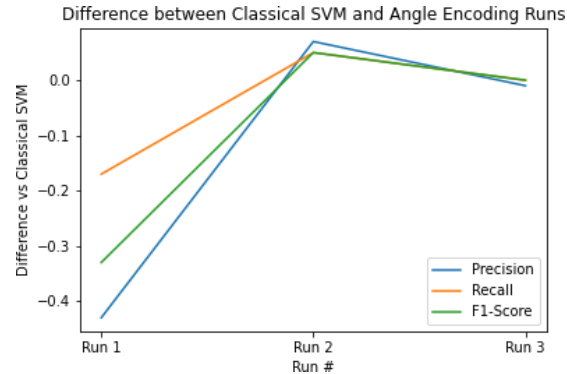